\documentclass{emulateapj}
\slugcomment{Submitted to \textsc{the Astrophysical Journal}}
\shorttitle{Effect of He Sedimentation on X-Ray Clusters}
\shortauthors{Peng \& Nagai}
\newcommand{\unitspace}{\ensuremath{\,}}
\newcommand{\usp}{\unitspace}
\newcommand{\numberspace}{\ensuremath{\;}}
\newcommand{\nsp}{\numberspace}

\newcommand{\kB}{\ensuremath{k_\mathrm{B}}} 
\newcommand{\mb}{\ensuremath{m_\mathrm{p}}} 

\newcommand{\beq}{\begin{equation}}
\newcommand{\eeq}{\end{equation}}

\newcommand{\gas}{_{\mathrm{gas}}}
\newcommand{\tot}{_{\mathrm{tot}}}
\newcommand{\he}{_{\mathrm{He}}}
\newcommand{\hy}{_{\mathrm{p}}}

\begin{document}

\title{Effect of Helium Sedimentation on X-ray Measurements of Galaxy Clusters}

\author{Fang Peng \altaffilmark{1}, Daisuke Nagai \altaffilmark{1,2,3}}
\altaffiltext{1}{Theoretical Astrophysics, California Institute of
Technology, Mail Code 130-33, Pasadena, CA 91125; fpeng@caltech.edu}
\altaffiltext{2}{Department of Physics, Yale University, New Haven, CT 06520}
\altaffiltext{3}{Yale Center for Astronomy \& Astrophysics, Yale University, New Haven, CT 06520; daisuke.nagai@yale.edu}

\begin{abstract}
The uniformity of the helium-to-hydrogen abundance ratio in X-ray emitting 
intracluster medium (ICM) is one of the commonly adopted assumptions in 
X-ray analyses of galaxy clusters and cosmological constraints derived
from these measurements.  In this work, we investigate the effect of He 
sedimentation on X-ray measurements of galaxy clusters in order to assess 
this assumption and associated systematic uncertainties.  
By solving a set of flow equations for 
a H-He plasma, we show that the helium-to-hydrogen mass ratio is significantly 
enhanced in the inner regions of clusters.  The effect of He sedimentation, if
not accounted for, introduces systematic biases in observable properties of 
clusters derived using X-ray observations.  
We show that these biases also introduce an apparent evolution in the observed 
gas mass fractions of X-ray luminous, dynamically relaxed clusters and hence 
biases in observational constraints on the dark energy equation of state 
parameter, $w$, derived from the cluster distance-redshift relation.  The Hubble 
parameter derived from the combination of X-ray and Sunyaev-Zel'dovich effect 
(SZE) measurements is affected by the He sedimentation process as well.  
Future measurements aiming to constrain $w$ or $H_0$ to better than 10\% 
may need to take into account the effect of He sedimentation.  We propose that 
the evolution of gas mass fraction in the inner regions of clusters should provide 
unique observational diagnostics of the He sedimentation process.

\end{abstract}


\section{Introduction}

Clusters of galaxies are powerful cosmological probes and have the potential to 
constrain properties of dark energy and dark matter.  Recent development in 
X-ray observations of galaxy clusters have produced a large statistical sample 
of clusters and start to deliver powerful cosmological constraints 
\citep{Allen2004Constraints_on_,Allen2008Improved,Mantz2007New_constraints,
Vikhlinin2008} that are complimentary to and competitive with other techniques 
(e.g., supernova, baryon acoustic oscillation, and weak lensing).
This has motivated construction of the next-generation of X-ray satellite missions 
(e.g., \emph{eROSITA}) to push the precision cosmological measurements based 
on large X-ray cluster surveys.  However, in the era of precision cosmology, the 
use of clusters as sensitive cosmological probes require solid understanding of 
cluster gas physics, testing of simplifying assumptions, and assessing associated 
systematic uncertainties.

One of the commonly adopted assumptions in X-ray cluster analyses include 
the uniformity of the helium-to-hydrogen abundance ratio with nearly primordial 
composition in X-ray emitting intracluster medium (ICM).  At present, there is 
no observational test of this assumption, since both H and He in the ICM are fully 
ionized, which makes it difficult to measure their abundances using traditional 
spectroscopic techniques.  Theoretically, on the other hand, it has long been 
suggested that heavier He nuclei slowly settle in the potential well of galaxy clusters 
and cause a concentration of He toward their center \citep{Abramopoulos1981On_the_equilibr,Gilfanov1984Intracluster,Qin2000BARYON-DISTRIBU,Chuzhoy2003Gravitational,Chuzhoy2004Element,Ettori2006Effects}. 
In the era of precision cosmology, this could be a source of significant systematic 
uncertainties in X-ray measurements of galaxy clusters and cosmological
parameter derived from these measurements \citep{markevitch2007Helium_abundance}.  

Thus, the primary goal of the present work is to assess the validity of this assumption 
and associated systematic uncertainties in X-ray measurements of key cluster 
properties as well as cosmological parameters derived from 
these observations.  In this work, we investigate the effects of He sedimentation 
on X-ray measurements of galaxy clusters by solving a set of diffusion equations 
for a H-He plasma in the ICM.  By taking into account observed temperature 
profiles obtained by recent X-ray observations
\citep{Vikhlinin2006Chandra_Sample,2007Pratt, 2008Leccardi,2008George}, 
we show that the observed temperature drop in the cluster outskirts lead to 
a significant suppression of He sedimentation, compared to the results based
on the isothermal cluster model \citep{Chuzhoy2004Element}.  Our analysis 
indicates that the He sedimentation has negligible effect on X-ray 
measurements in the outer regions of clusters (e.g., $r_{500}$), and it does
not affect cluster mass measurements obtained at the sufficiently large
cluster radius.  The effect of He sedimentation, on the other hand, 
introduces increasingly larger biases in X-ray measurements in the inner regions 
and could affect cosmological constraints, including the dark energy equation of 
state parameter $w$ derived from distance-redshift relation as well as $H_0$ 
derived from the combination of X-ray and Sunyaev-Zel'dovich effect (SZE). 
 
The paper is organized as follows.  In \S~\ref{sec:xray}, we describe the 
dependence of X-ray clusters measurements on 
He abundance 
in the ICM. The physics of He sedimentation and cluster models are discussed in 
\S~\ref{sec:diffusion}.  In \S~\ref{sec:results}, we present results of our He 
sedimentation calculations and investigate their effects on cluster properties and 
cosmological constraints derived from X-ray cluster observations.  Main conclusions 
are summarized in \S~\ref{sec:conclusions}.

\section{Effect of He Abundance on X-ray Measurements} 
\label{sec:xray}

The observed X-ray surface brightness of a cluster is primarily from 
bremsstrahlung continuum emission of electrons scattering off of 
protons and He nuclei, which is given by the integral of the emission 
along the line of sight,
\begin{eqnarray}
\label{S_X.e}
S_X & \propto & \int dl \, (n_e n\hy \Lambda_{e {\rm p}} + n_e n\he \Lambda_{e\rm{He}}) \\
       &\propto  & n\hy^2  (1+2x)(1 + 4x)  \Lambda_{e {\rm p}} \ ,
       \label{sx_x.e}
\end{eqnarray}
where $x\equiv n\he/n\hy$ is the 
He-to-H abundance ratio 
and $\Lambda_{ei}$ is the band-limited cooling function resulted
from free-free emission of electrons scattering off ion species $i$,
which is proportional to $Z_i^2$.  The derivation of the second
expression takes into account the charge dependence of cooling function
and the fact that the number density of electron is given by $n_e = n\hy + 2 n\he = n\hy
(1+2x)$.  Since $S_X$ is observed and fixed, the proton number density
$n\hy$ inferred from X-ray observations depends on x, as 
$n\hy \propto 1/\sqrt{(1+2x)(1+4x)}$.  The derived gas mass density 
therefore depends on x as,
\beq
\label{rgas.e}
\rho\gas \propto n\hy + 4 n\he \propto \left(\frac{1+4x}{1+2x}\right)^{1/2} \ .
\eeq
The gas mass $M\gas$ is the volume integral of equation~(\ref{rgas.e}). 
The hydrostatic mass profile of a spherically-symmetric cluster depends 
on the local value of He abundance through the mean molecular weight 
of particles $\mu$, 
\beq \label{mhse}
M\tot(<r) \propto \frac{r T_e}{\mu} \frac{d{\rm log} \rho_{\rm gas}}{d{\rm log} r} 
\propto \frac{1}{\mu} = \frac{2 + 3x}{1 + 4x} \ .
\eeq
Note that the hydrostatic mass at a fixed mean overdensity, $\Delta$, 
is given by $M_{\Delta} \propto (T_e/\mu)^{3/2}$.  The gas mass fraction
is then defined as $f\gas \equiv M_{\rm gas}/M_{\rm tot}$.   

For the primordial abundance (X = 0.75, Y = 0.25 by mass), the He-to-H 
abundance ratio is $x=0.083$ and the mean molecular weight is $\mu = 0.59$.
If, for example, the He-to-H abundance ratio is enhanced by a factor of 
two from the 
primordial value, the analysis based on the assumption of the 
primordial abundance causes an underestimate of $\rho_{\rm gas}$ 
by 5\% and an overestimate of $M_{\rm tot}$ by 12\%.  

\section{Helium Sedimentation in X-ray Custers} \label{sec:diffusion}

\subsection{Diffusion Equations}\label{sec:diffusioneqn} 

Particle diffusion in clusters is characterized by the Burgers equations 
of a multicomponent fluid \citep{burgers69:composite_gases}.  
Each species $s$ obeys an equation of continuity and momentum 
conservation,
\begin{eqnarray}
  \frac{\partial n_s}{\partial t}+ \frac{1}{r^2}\frac{\partial (r^2n_s u_s)}{\partial r} &=&0~
  \label{continuity.e}\ ,\\
  \frac{\partial P_s}{\partial r} + n_s A_s \mb g - n_s Z_s eE &=& \sum_t K_{st} (w_t - w_s) \ .
  \label{burgers.e}
\end{eqnarray}
Here the species $s$ has mass $A_s \mb$, 
charge $Z_s e$, density $n_s$, partial pressure $P_s$, and velocity
$u_s$, 
where $\mb$ is the proton mass.
The center-of-mass of a fluid element moves with a velocity $u
= \sum_s n_s A_s u_s/\sum_s n_s A_s$.  The differential, or diffusion,
velocity between species $s$ and the fluid element is then $w_s = u_s
- u$. These diffusion velocities satisfy mass and charge conservation,
\begin{eqnarray}
  \label{mc.e}
  \sum_s A_s n_s w_s &=& 0 \ , \\
  \label{cc.e}
  \sum_s Z_s n_s w_s &=& 0 \ .
\end{eqnarray}
Note that the summations include both ions and electrons. To satisfy 
these conservation laws, for each sinking helium nuclei, there are roughly
four protons and two electrons that float up.

Equation~(\ref{burgers.e}) describes forces acting on a species $s$, and 
it is the balance of these forces that ultimately determines the rate of 
sedimentation.  For a sinking He nucleus, the gravitational force ($g$) is 
counteracted by three types of forces provided by the induced electric field 
($E$), the pressure gradient ($dP_s/dr$) of the ICM, and the drag force due 
to collisions with surrounding particles. Note that we neglected small terms including the 
inertial term (${\rm d}u_s/{\rm d}t$) and the shear stresses (or viscosity) 
due to collisions among the same species.   
We also neglected terms related to the coupling of thermal and particle diffusions, 
which lead to an underestimate of the He-to-H mass ratio by $\lesssim$~20\% 
in the radial range considered in this work (see \S~\ref{sec:discussions}).

Note that the sedimentation destroys hydrostatic equilibrium since 
redistribution of particles introduce a temporal change in the total
gas pressure.  However, hydrostatic equilibrium can be 
restored quickly.  This equilibrium restoring acquires a net inflow 
with a mean velocity $u$,
\beq \frac{du}{dt} = - \frac{1}{\rho\gas} \frac{\partial
P\gas}{\partial r} - g \ ,
\label{du.e}
\eeq
where $P\gas = \sum_s n_s k_{\mathrm B} T$ is the total gas pressure
for ideal gas, and $\rho\gas = \sum_s n_s A_s m_u$ is the gas density.
Equations~(\ref{continuity.e})-(\ref{du.e}) describe the process of
particle diffusion in the ICM.

\subsection{Resistance Coefficient} \label{sec:resistance}

When the plasma is sufficiently rarefied, which is the case for the ICM, 
particle pairs interact via a pure Coulomb potential.  In the absence of 
magnetic field and turbulence, the resistance coefficient is given by 
\citet{chapman.cowling} as,
\begin{equation}
 K_{st}^{B=0} \cong \frac{4 \sqrt{2 \pi}}{3} 
 \frac{e^4Z_s^2Z_t^2 \mu_{st}^{1/2}}{(\kB T)^{3/2}} n_s n_t \ln \Lambda_{st}  \ ,
\label{resist.e}
\end{equation}
where $\mu_{st} = A_sA_t\mb/(A_s+ A_t)$ is the reduced mass of species
$s$ and $t$, and a typical value of the Coulomb logarithm is 
$\ln \Lambda_{st} \simeq 40$ for the H-He plasma in the ICM.  Resistance coefficient 
describes the momentum transfer rate of species $s$ per unit volume due to 
collisions with species $t$ (in units of $\mathrm{g\,cm^{-3}\,s^{-1}}$).
The resistance coefficient, $K_{st}\propto T^{-3/2}$, is inversely 
proportional to the ICM temperature. Therefore,
particle transport is more efficient in the ICM with higher 
temperature. 
Note that heat transport via the same particle collision physics gives that 
the thermal conductivity depends on $T$ as $\kappa\propto T^{5/2}$. 

However, magnetic field and turbulence, present in real clusters, can 
significantly modify the resistance coefficients.  To date, theoretical work on 
this subject has predicted a wide range of the magnetic suppression factor, 
$f_B \equiv \kappa/\kappa_{\rm Sp} \sim 0.1-1$, where $\kappa_{\rm Sp}$ is 
the Spitzer thermal conductivity \citep{Spitzer1962}, depending sensitively 
on the strength and the geometry of magnetic field as well as the nature of 
MHD turbulence.   In the case of a tangled magnetic field, the thermal conductivity 
may be moderately suppressed ($f_B \sim 0.1-0.2$) relative to the Spitzer value 
\citep{Narayan2001THERMAL-CONDUCT,Chandran2004Thermal_Conduc}. 
Magnetothermal instability tends to drive magnetic field lines to a radial 
direction, which gives the suppression of $f_B\gtrsim 0.4$ 
\citep{Parrish2008}.  It was also suggested that MHD turbulence may 
provide the magnetic suppression factor of order unity 
\citep{Cho2004,Lazarian2007}.  

Observationally, one can use the strength of observed temperature gradients 
in the ICM to constrain the efficiency of thermal conductivity.  Observed 
large-scale, negative ICM temperature gradients set the upper limit on the 
thermal conductivity, $f_B \lesssim 0.2$ for a $T_X=10$~keV cluster with 
a typical age of 7~Gyr \citep{Loeb2002}.  
Observations of A754 also yield
a suppression factor $f_B < 0.1$ for the bulk of the ICM \citep{markevitch03}.
The width of cold fronts observed 
by recent X-ray observations suggests a considerably smaller suppression 
factor ($f_B \ll 0.1$) \citep[][for a review]{Markevitch2007_ColdFront}.  
This, however, is a local constraint which is not applicable to a cluster as 
a whole.  
One might also imagine that the abundance profile of heavy nuclei 
(e.g., Fe) with X-ray emission lines may provide further insights into the 
efficiency of particle diffusion.  Though possible in principle, interpretation 
of the abundance profile is complicated by the fact that the ICM is enriched 
continuously by stripping of metal-enriched gas from cluster galaxies, and 
that the process of sedimentation is generally slower for heavier nuclei. 

Given the current large uncertainties in diffusion coefficients due to the 
lack of knowledge on magnetic field and turbulence in the ICM, we 
parametrize the effective resistance coefficients as 
$K_{st} = f_B^{-1} K_{st}^{B=0}$, where we take $f_B$ as a free parameter.
In order to illustrate how our results depend on the suppression factor, we 
consider two cases with $f_B = 1$ (un-magnetized) and $0.2$ (tangled 
magnetic field).  The former should be taken as an extreme model, while 
the latter is roughly corresponds to the current observational limit discussed 
above.
Note, however, that all current observations are consistent with $f_B = 0$ 
(conduction fully suppressed).  We, therefore, caution that $f_B$ may 
be orders of magnitude below the unity. 

\subsection{Sedimentation Velocity} \label{sec:drift_steady}

To develop physical insights into the process of He sedimentation, it
is useful to consider a drift velocity of a trace He particle in a background of hydrogen, 
i.e, $n\hy \gg n\he$ 
and $w\hy = 0$.  In this limiting case, equations~(\ref{burgers.e}) for 
the two species are decoupled.  The right-hand side of the equations 
for H vanishes, thereby fixing an electric field, 
$eE = 0.5\,\mb g$.  Substituting $E$ into the equation 
of motion for He, we obtain the sedimentation velocity 
of He nuclei as $w\he = 3\,\mb g n\he /K_{\rm{pHe}}$,
which gives
\begin{eqnarray}
w\he &\simeq & 80\usp{\rm km\,s^{-1}} f_B \nonumber \\ 
          &             & \times \left(\frac{T}{10\usp{\rm keV}}\right)^{3/2} 
\left(\frac{g}{10^{-7.5}\usp{\rm cm\,s^{-2}}}\right) \left(\frac{n\hy}{10^{-3}{\rm cm^{-3}}}\right)^{-1} \ ,  
\label{vsed.e}
\end{eqnarray}
where the induced electric field counteracts gravity and suppresses
the sedimentation speed by $25\%$.  At a fixed 
density, the sedimention velocity is generally larger for higher temperature 
($T$) and gravity ($g$).  Note that the pressure gradient ($dP/dr$) term in 
equation~(\ref{burgers.e}) further suppresses the sedimentation process.  
Typical sedimentation timescale in clusters is generally longer than 
the Hubble time.  
The equilibrium distributions to the Burgers equation
\citep{Abramopoulos1981On_the_equilibr,Qin2000BARYON-DISTRIBU,
Chuzhoy2003Gravitational} are therefore not applicable for clusters.  
Instead, a full time-dependent calculation is required for this analysis.

\subsection{Cluster Models and Initial Conditions}\label{sec:cluster}

We set up cluster models and initial conditions as follows. Initially, 
we assume that the ICM consists of a primordial H and He plasma 
uniformly throughout clusters.  We ignore the contribution of heavier 
elements of $Z >2$. 

For the total mass distribution, we adopt the Navarro-Frenk-White
(NFW) density profile \citep{Navarro1997A_Universal_Den},
\beq
\rho(r) = \frac{\rho_s}{x'(1+x')^2}\ ,
\label{rho_NFW.ee}
\eeq
where $x' \equiv r/r_s$,
$r_s$ is a scale radius, and $\rho_s$ is a normalization constant.
Mass enclosed within a radius $r$ is then given by 
$M(x') = 4\pi\rho_sr_s^3 \left[\ln(1+x') - x'/(1+x')\right]$.
Throughout this work, we define the cluster mass to be $M_{\Delta} =
(4\pi/3)r_{\Delta}^3\Delta\, \rho_{crit}$, where $r_{\Delta}$ is a radius
of a spherical region within which the mean enclosed mass density is
$\Delta$ times the critical density of the universe $\rho_{crit}$.  We 
adopt $\Delta=500$ and the concentration 
$c_{500} \equiv r_{500}/r_s = 4$ (V06). We also
consider $\Delta=2500$, where some of the X-ray measurements are also
made. In this cluster model, $r_{2500}/r_{500} = 0.46$.

\begin{figure}[t]
  \centering
  \includegraphics[clip=true,trim=0.0cm 2.0cm 0.0cm 3.0cm,width=3.8in]{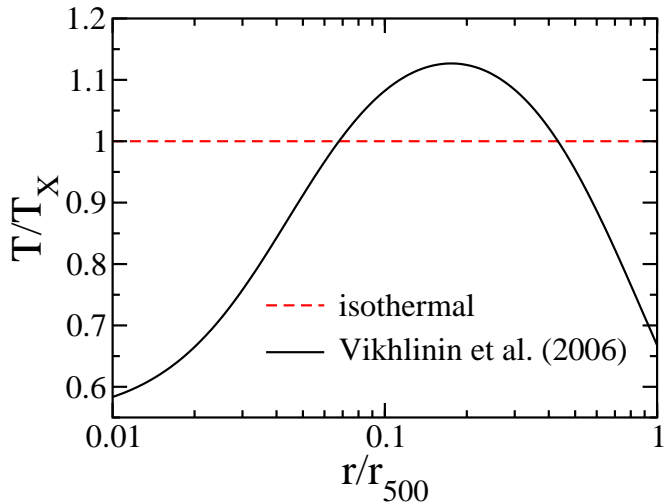}
  \caption{Temperature profiles of X-ray clusters used in our models:
isothermal temperature profile (\emph{dashed line}) and observed
temperature profile obtained by \citet{ Vikhlinin2006Chandra_Sample}
(\emph{solid line}).  }
  \label{temp.f}
\end{figure}

We consider two ICM temperature models: (1) the isothermal temperature 
profile $T(r) = T_{X} $, and (2) the observed temperature profile obtained 
with deep Chandra observations of nearby relaxed clusters (V06) given by
\beq
\frac{T(r)}{T_X} = 1.216~\frac{(\tilde{x}/0.045)^{1.9}+0.45}{(\tilde{x}/0.045)^{1.9}+1}\frac{1}
{\left[1+(\tilde{x}/0.6)^2\right]^{0.45}} \ ,
\label{temp.ee}
\eeq 
where $\tilde{x}=r/r_{500}$, and $T_X$ is the X-ray spectral
temperature, which is related to $M_{500} = M_5 (T_X/5\nsp {\rm
keV})^\alpha$, where $M_5 = 2.89\times10^{14}\nsp h^{-1}\usp M_\odot$
and $\alpha = 1.58$ (V06).  For an illustration, we plot these ICM
temperature profiles in Figure~\ref{temp.f}. The observed temperature
peaks around $r/ r_{500} \simeq 0.2$, and decreases at both inner
and outer radii. For example, the temperature drops by nearly a factor
of $2$ from the peak value at $r/r_{500} = 0.01$ and $1$.

We set up the initial gas distribution by assuming hydrostatic
equilibrium of the ICM in the potential well of clusters dominated 
by dark matter,
\beq
\frac{k_{\mathrm B}} {\mu m_u} \frac{d (\rho\gas T)}{dr} = -\rho\gas g(r) \ ,
\label{static.e}
\eeq
where the gravitational acceleration at a radius $r$ is $g(r)\equiv
GM(r)/r^2$.  For each temperature profile $T(r)$, we derive the initial 
gas density profile as,
\begin{eqnarray}
\rho\gas (r) & = & \rho_{\mathrm{gas},0} \left(\frac{T_0}{T(r)}\right) 
\exp \left(-\frac{\mu m\hy}{\kB}\int_0^r \frac{g(r')}{T(r')} dr' \right)  \ ,        
\label{rho_gas}
\end{eqnarray}
where the central density $\rho_{\mathrm{gas},0}$ is normalized 
by requiring $M_{\rm gas}/M (r_{500}) \equiv f_{\rm gas,500}$ = 0.15 (V06),
and $T_0$ is the central temperature for each ICM temperature model.
For isothermal case, the density profile reproduces the analytical form 
given by $\rho\gas (r) = \rho_{\mathrm{gas},0} \exp(-\eta \mu)
(1+x')^{\eta \mu/x'}$ with $\eta = 4\pi G m\hy \rho_s r_s^2/(\kB T_X)$ \citep{Makino1998X-RAY-GAS-DENSI}.

Finally, we consider two types of mass accretion histories (MAHs) of clusters:
(1) a static cluster with a fixed mass, (2) the MAHs of cluster-size halos calibrated 
with N-body simulations of cluster formation.  For the latter, we adopt an analytical
expression of averaged MAHs of halos given by 
\citet{2002vandenbosch_the_universal},
\beq \log 
\left(\frac{M}{M_0}\right) = -0.301
\left[\frac{\log(1+z)}{\log(1+z_f)}\right]^\nu \ ,
\label{massah.e}
\eeq
where $M_0$ is the cluster mass at present time, $z_f$ is the formation 
redshift defined as $M(z_f)/M_0=0.5$, and $\nu$ is a parameter that 
determines the shape of MAHs of dark matter halos, and it is 
strongly correlated with $z_f$.  This functional form provides a reasonable 
description of the MAHs of simulated clusters for a typical value of $\nu$ in the 
range of $1.4-2.3$ \citep{2004TasitsiomiDensity} and an average formation 
redshift of $\langle z_f \rangle \simeq 0.6$ 
\citep[][see also Berrier et al. 2008]{2005Cohn}. 
\footnote{Note that the equation~(\ref{massah.e}) provides a reasonable 
description of the MAHs of simulated galactic-size halos \citep{Wechsler2002} 
by tuning a free parameter $f$ to $0.254$ 
\citep[see Appendix A in][]{2002vandenbosch_the_universal}.  For the cluster-size
halos, we find that $f=0.656$ provides a reasonable description of their MAHs 
and the average formation time of a large statistical samples of clusters-size halos 
extracted from N-body simulations \citep[][see also Berrier et al. 2008]{2005Cohn}.}  
The points that are particularly relevant for this work are that a typical 
present-day cluster grows by a factor of two in mass in the past 
$\approx 6$~Gyr since $z \approx 0.6$, 
while an equal-mass cluster at higher redshift ($z=1$) forms in a considerably 
shorter timescale ($\approx 1-2$~Gyr).  This suggests that the high-z clusters 
are generally much less affected by the process of He sedimentation than their 
low-redshift counterparts.  
In \S 4.1-4.3, we use the simple static cluster model to investigate the 
dependence of the He sedimentation efficiency on some of the key 
cluster parameters, including the temperature, the age, and the 
magnetic suppression factor. 
In \S~\ref{sec:fgasEvol}, we use the realistic MAHs 
to investigate evolution in X-ray observable properties with redshift.  
Throughout this work, we use cosmological parameters: 
$\Omega_{\rm M} = 0.3$, $\Omega_\Lambda = 0.7$, 
$\Omega_{\rm b} = 0.0462$, and $h = 0.7$ \citep{2008Komatsu_Five_Year}. 

\section{Results} \label{sec:results}

\subsection{Spatial Distribution and Evolution of He Abundance}

Starting from the initial cluster model described in \S~\ref{sec:cluster}, 
we follow the diffusion process in the H and He plasma by solving
equations~(\ref{continuity.e})-(\ref{du.e}) numerically.   At each time step, 
equations (\ref{burgers.e})-(\ref{cc.e}) are solved to obtain diffusion 
velocities ($w_s$) for H, He and electrons as well as electric field ($E$).  Using 
these diffusion velocities, 
we update the abundance of each species by solving
equations~(\ref{continuity.e}) and (\ref{du.e}) together.
We repeat this procedure through the mass accretion histories of galaxy clusters.  
We use 600 spatial grids logarithmically spaced in a computational 
domain of $10^{-3} \leq r/r_{500} \leq 10$ and set the diffusion velocities 
to zero at both inner and outer boundaries.  These choices ensure that 
results are robust in regions of our interest ($0.01 < r/r_{500} \leq 1$).
\begin{figure}[t]
  \centering
  \includegraphics[clip=true,trim=0.0cm 2.0cm 0.0cm 3.0cm,width=3.8in]{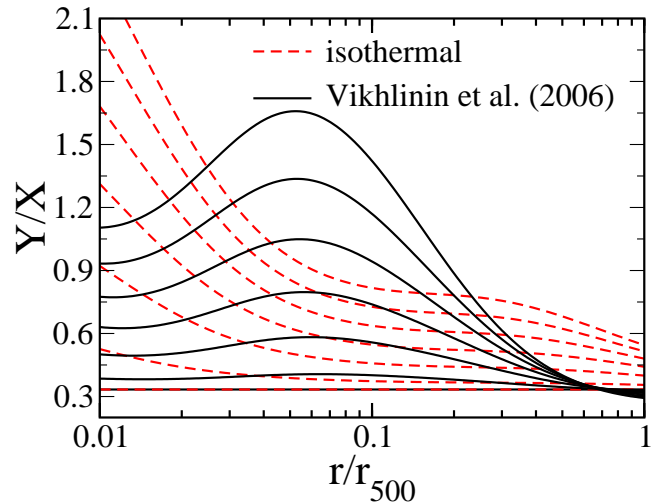}
  \caption{The spatial distribution of helium-to-hydrogen mass
fraction ($Y/X$) in a $T_X=10$~keV static cluster with $f_B=1$.  Lines indicate
two types of ICM temperature profiles shown in Fig.~\ref{temp.f}. From 
bottom to the top, the set of curves shown are for cluster ages of 0, 
1, 3, 5, 7, 9, and 11~Gyr, respectively.}
  \label{YXratio.f}
\end{figure}

Figure~\ref{YXratio.f} shows evolution of a spatial distribution of the 
He-to-H mass fraction ($Y/X$) of a $T_X=10$~keV cluster with $f_B=1$.
Here we compare results of the isothermal model and those of the V06 
model.  
This comparison shows that the observed temperature drop in the outer regions 
significantly suppresses the process of He sedimentation in the outskirts ($r/r_{500} > 0.3$)
of clusters. The efficiency of He sedimentation, however, is enhanced in the inner regions 
near the peak of the observed temperature profile, and it is suppressed again in the inner 
most regions ($r/r_{500} < 0.02$) due to the observed temperature drop at these radii.
At $r=r_{500}$, the He 
abundance in the V06 model is very close to the primordial value 
(Y/X = 0.333). The effect of He sedimentation is negligible at this radius,
and these results are fairly independent of cluster age and the value of 
$f_B$.   These conclusions are in stark contrast with the result of 
the isothermal model which gives considerably larger He abundance 
of $Y/X=0.479$.  In the V06 model, the value of $Y/X$ increases rapidly 
toward inner regions and peaks at $r= 0.06\nsp r_{500}$.  The mass 
fraction ratio increases to $Y/X=0.405$ at $r=r_{2500}$ and $1.0$ at 
$r=0.1\nsp r_{500}$ for the cluster age of 7~Gyr.

\begin{figure}[t]
  \centering 
  \includegraphics[clip=true, trim=0cm 0cm 0cm 1cm,width=3.5in]{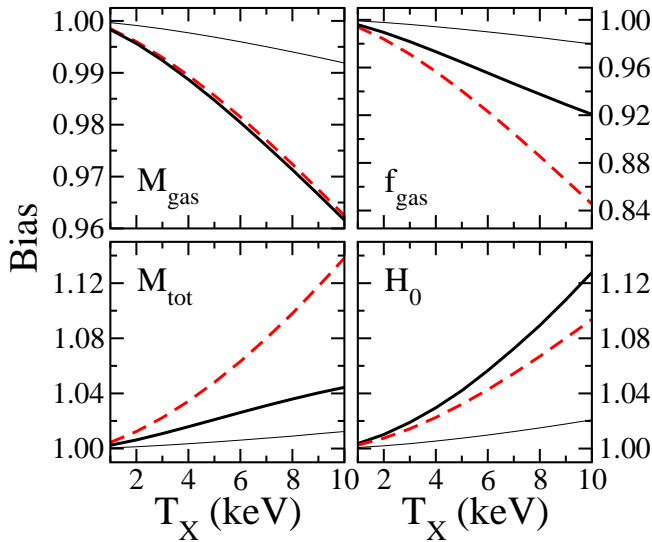}
  \caption{Biases in X-ray measurements of galaxy clusters at 
  $r_{2500}$ as a function of $T_X$ for a static cluster age of $7\usp{\rm Gyr}$.  
  Lines indicate the isothermal model (\emph{dashed line}) and
  the V06 model with $f_B=1$ (\emph{thick-solid line}) and $f_B=0.2$ 
  ({thin-solid line}).  }
  \label{bias.f}
\end{figure}

\subsection{Effect on Gas Mass, Total Mass, and Gas Mass Fraction}

As we discussed in \S~\ref{sec:xray}, the spatial variation in the He abundance 
caused by the mass segregation could introduce observational biases in 
X-ray measurements of galaxy clusters.  Figure~\ref{bias.f} shows biases in 
X-ray measurements of gas mass, total mass, and gas mass fraction at 
$r_{2500}$ for three sedimentation models: the isothermal model with 
$f_B=1$ ({\it dashed} line), the V06 model with $f_B=1$ ({\it thick-solid} line) 
and $f_B=5$ ({\it thin-solid} line).  Results are shown for a typical cluster 
age of $7$~Gyr and for a range of average cluster temperatures 
($T_X = 1-10$~keV).  Here we define these biases to be ratios between the 
quantities derived by assuming the primordial abundance and their true values 
in our cluster models, which corresponds to the values that would have been 
obtained if the sedimentation effect has been taken into account in X-ray 
data analyses.  

Comparing results of two different models, we find that biases in X-ray 
measurements of total cluster mass at $r_{2500}$ are considerably smaller 
in the V06 model.  In the case of a hot ($T_X=10$~keV), un-magnetized 
($f_B=1$) cluster, for example, the total mass is overestimated by 4\% and 
14\% for the V06 model and isothermal model, respectively.  The biases in $M\gas$, 
on the other hand, are underestimated by 4\% for both models.  Note that these
values are similar for two different temperature models.  This is because $M\gas$ 
measurements, obtained by integrating the gas density over the cluster volume, 
depend primarily on the average ICM temperature ($T_X$), and relatively 
insensitive to the details of ICM temperature profiles.  $f\gas$ is then biased low 
by 8\% and 16\% for V06 model and isothermal model, respectively. 
Note that the bias in $f_{\rm gas}$ above is obtained by integrating from 
the cluster center to the radius $r$.  This yields a large bias than the one 
evaluated by using the local He abundance at $r$ 
\citep{markevitch2007Helium_abundance}.
\begin{figure}[t]
  \centering
  \includegraphics[clip=true, trim=0cm 0cm 0cm 1cm,width=3.5in]{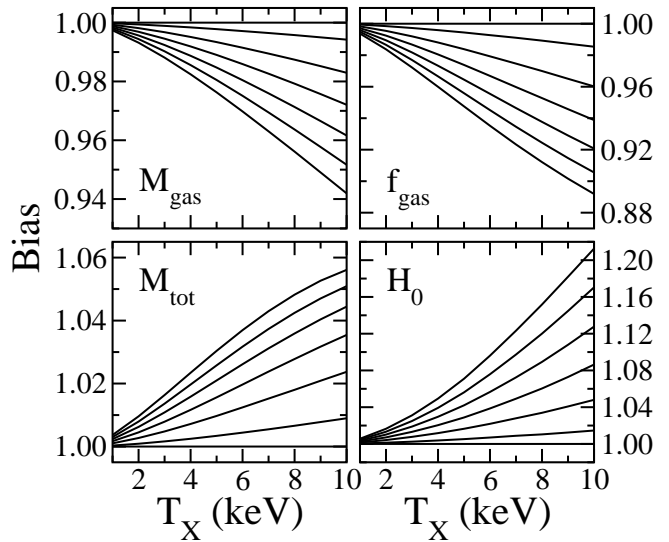}
  \caption{Biases in X-ray measurements of galaxy clusters $r_{2500}$ 
  as a function of cluster age. The curves starting from the flat line 
  corresponds to cluster ages of 0, 1,  3, 5, 7, 9, and 11~Gyr.  Results are 
  shown for static clusters of the V06 model with $f_B=1$.}
  \label{bias_t.f}
\end{figure}

The biases in X-ray measurements are very sensitive to the average ICM 
temperature ($T_X$) as well as the magnetic suppression factor ($f_B$). 
The biases are considerably smaller in cluster with smaller values of $T_X$ 
or $f_B$, because the sedimentation process is slow in these clusters with 
larger resistance coefficient.  We find that the biases at $r_{2500}$ become 
negligible ($\lesssim 2\%$) for $T_X \lesssim 3$ keV or $f_B\lesssim 0.2$.  
These biases also depend sensitively on the age of clusters.  Figure \ref{bias_t.f} 
illustrates that these biases increase rapidly with the cluster age and cause 
the apparent evolution of X-ray measurements of hot, dynamically relaxed 
clusters.  Our sedimentation model based on the observed V06 temperature 
profile predicts that the biases in $f\gas$ are likely less than 10\% for a realistic 
range of cluster parameters (including the cluster age, $T_X$, and $f_B$).  
Note also that the biases become negligible in cluster outskirts ($r\sim r_{500}$) 
for all systems.  

\subsection{X-ray and SZE-derived Hubble Constant} \label{sec:hubble}

Measurements of angular diameter distance derived from the combination
of X-ray and Sunyaev-Zel'dovich Effect (SZE) observations also depend 
on the assumed He abundance, x, as,
\beq 
d_{\rm A} \propto \frac{y^2}{S_X T_e^2} \frac{1+4x}{1+2x} \ ,
\label{da.e}
\eeq 
where the expression was obtained by canceling electron densities
in both the X-ray surface brightness $S_X \propto n_e^2 d_A
(1+4x)/(1+2x)$ (see eq.~[\ref{sx_x.e}]) and the SZE comptonization
parameter $y \propto n_e T_e d_A$.  Since $H_0 \propto d_A^{-1}$, the
X-ray+SZE derived $H_0$ measurements could be affected by the
He sedimentation process.  As shown in the bottom-right panels of 
Figures \ref{bias.f} and \ref{bias_t.f}, our model suggests that the 
sedimentation could introduce biases in the X-ray+SZE derived $H_0$ 
measurements high by about $\lesssim 15\%$, 
with the exact value depending on the cluster age, temperature, as well as 
the magnetic suppression factor. 
This effect is of the same sign and order as the 6\% offset seen between
the X-ray and SZE-derived Hubble constant
($H_0 = 77.6^{+4.8}_{-4.3}\usp {\rm km \,s^{-1}\,Mpc^{-1}}$) 
\citep{Bonamente2006Determination} and result of the Hubble Key Project 
($H_0 = 73 \usp {\rm km \,s^{-1}\,Mpc^{-1}}$) \citep{Freedman2001}. 
Given the errors in the current X-ray+SZE $H_0$ measurements, our 
models with $f_B=1$ and $0.2$ are both consistent with the observed offset.
As proposed by \citet{markevitch2007Helium_abundance}, the comparison
of the X-ray+SZE derived $H_0$ and independent $H_0$ measurements 
could be used to constrain the He abundance in clusters.

\begin{figure}[t]
  \centering \includegraphics[clip=true,trim=0.2cm 0.5cm 0.0cm 3.5cm,width=3.6in]{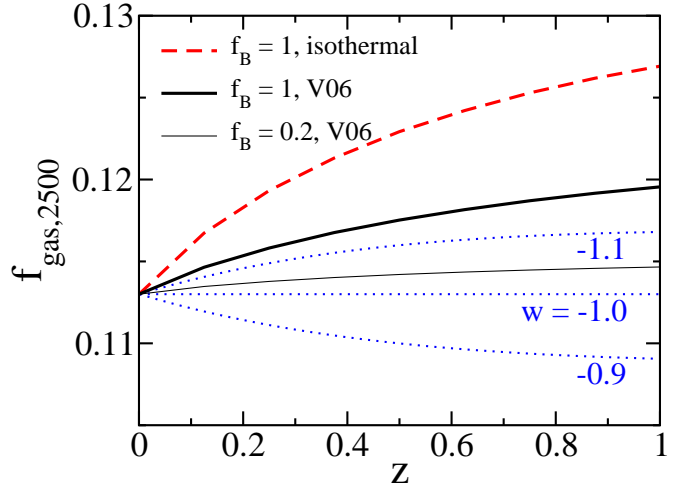}
  \caption{Evolution of cluster gas mass fraction enclosed within $r_{2500}$
  as a function of redshift.  Results are for a 10~keV cluster with realistic MAHs 
  given by eq.~(\ref{massah.e}).  Lines indicate the isothermal model with 
  $f_B=1$ (\emph{dashed line}), V06 model with $f_B=1$ (\emph{thick-solid line}) 
  and $f_B=0.2$ (\emph{thin-solid line}).  \emph{Dotted} lines indicate
  the evolution in $f\gas$ caused by changes in the dark energy equation of 
  state $w$ by 10\% relative to the fiducial $\Lambda$CDM cosmological 
  model with no sedimentation (indicated with a straight dotted line).  
  }
  \label{fgas_zobs.f}
  \vspace{2mm}
\end{figure}

\subsection{Evolution of Cluster Gas Mass Fraction} \label{sec:fgasEvol}

The evolution of cluster gas mass fraction of X-ray luminous,
dynamically, relaxed clusters can provide powerful observational
constraints on the equation of state of dark energy $w$
\citep{1996Sasaki,1997Pen,2003Ettori,Allen2004Constraints_on_,2006LaRoque,Allen2008Improved}.
The sensitivity to $w$ lies in the dependence of the observed $f\gas$ to
the angular diameter distance, $d_{\rm A}(w,z)$,
\beq
f\gas(w,z) = 0.113
\left[\frac{d_{\rm A}(w,z)}{d_{\rm A}(w=-1.0,z)}\right]^{1.5} \ ,
\eeq
where the observed $f\gas$ is $0.113$ at $z=0$.  Recent 
measurements of the $f\gas$
evolution based on \emph{Chandra} X-ray observations of 42 bright,
dynamically relaxed galaxy clusters yielded $w = -1.14 \pm 0.31$, by
assuming a flat geometry and standard priors on $\Omega_b h^2$ and $h$. 
The combined analysis of $f\gas$ plus CMB and SNIa measurements 
constrains $w$ to better than $10\%$, $w = -0.98\pm 0.07$\citep{Allen2008Improved}.

Here we point out that the effect of He sedimentation can introduce an 
apparent evolution in X-ray measurements of cluster gas mass 
fractions, which could lead to systematic biases in observational constraints 
on the dark energy equation of state, $w$.   This is illustrated in 
Figure~\ref{fgas_zobs.f}, where we show evolution of the observed gas 
mass fraction at $r_{2500}$ by using the realistic, time-dependent MAHs
given by equation~(\ref{massah.e}) in order to follow the He sedimentation process 
in a growing cluster-size halo, starting from a proto-cluster at $z=5$.
Our sedimentation models show that the observed gas mass 
fraction is overestimated by $12\%$, $6\%$, and $2\%$ at $z=1$, for the 
isothermal model with $f_B =1$, V06 model with $f_B = 1$ and with $f_B = 0.2$, 
respectively.  The apparent evolution in $f\gas$ arises because clusters at 
lower-redshifts have had more time to experience the sedimentation on average.  

\begin{figure}[t]
  \centering \includegraphics[clip=true,trim=0.2cm 0.5cm 0.0cm 3.5cm,width=3.6in]{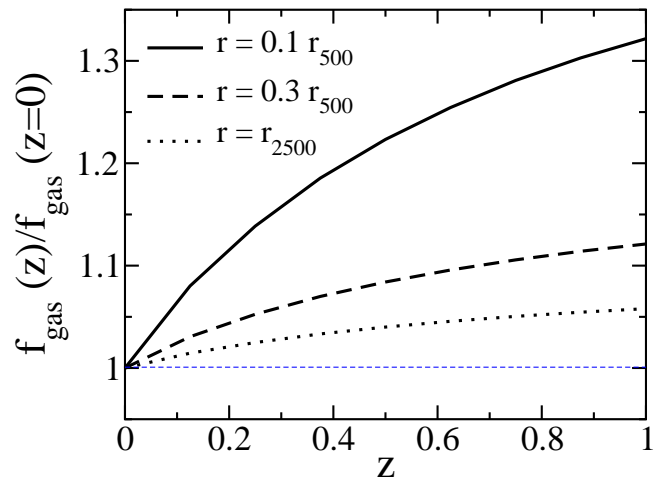}
  \caption{Evolution of cluster gas mass fraction enclosed within 
  three cluster-centric radii, $r=0.1\nsp r_{500}$ (\emph{solid lines}), 
  $r=0.3 \nsp r_{500}$ (\emph{dashed lines}), and $r=r_{2500}$ 
  (\emph{dotted lines}).  Results are for a 10 keV cluster with realistic MAHs
  given by eq.~(\ref{massah.e}), for the V06 model and $f_B=1$.  }
  \label{fgas_zobs_inner.f}
  \vspace{0.2mm}
\end{figure}

As shown in Figure~\ref{fgas_zobs.f}, the effect of He sedimentation is degenerate 
with changes in the dark energy equation of state, $w$.  Note that the changes 
in $w$ by -10\% (relative to the fiducial $\Lambda$CDM cosmology with $w=-1$) 
corresponds to the changes in $f\gas$ by about $3\%$ between $z=0$ and $1$.  
The effect of He sedimentation is to make the best-fit $w$ more negative, i.e., 
$w < -1$.  For example, results of the V06 model with $f_B=1$ ($0.2$) are 
degenerate with the cosmological models with $w=-1.18$ ($-1.04$) without 
sedimentation.  The isothermal model with $f_B=1$, which should be 
taken as an extreme case, requires $w=-1.38$.  Given the measurement 
uncertainties ($\approx 30\%$ in $w$), current constraints on $w$ should not
be affected significantly by the effect of He sedimentation.  However, future X-ray 
measurements aiming to constraint $w$ to better than 10\% will need to take into 
account of this effect.  Note that the effect would be larger for a population of 
dynamically relaxed clusters than the average population considered here.

Given that the variation in the He abundance in clusters are unknown at present, 
it would be interesting to ask whether the current X-ray data could be used to 
constrain the abundance distribution.  
Here we propose that the $f\gas$ evolution in the inner cluster regions, where
the effect of sedimentation is expected to be larger, can be used as sensitive 
probes of the He sedimentation process.  In Figure \ref{fgas_zobs_inner.f}, 
we illustrate evolution of $f\gas$ (V06 model with $f_B=1$) at three 
different radii, including $r=0.1\,r_{500}$, $0.3\,r_{500}$, and $\,r_{2500}$.   
This figure shows that changes in $f\gas$ between $z=0$ and 1 are 
amplified significantly in the inner regions of clusters.   
At $r=0.1\,r_{500}$, $f\gas$ evolves by $\sim 30\%$ from $z = 0$
to 1, which could be detected with current data.  If the gas mass fraction 
evolution is detected, it can provide unique constraints on the efficiency
of He sedimentation, which in turn provides constraints on the magnetic 
and turbulent suppression of particle diffusion in the weakly magnetized 
ICM.  Non-detection is also interesting, as it will provide an upper limit on 
the efficiency of particle diffusion in the ICM. 

\section{Discussions}
\label{sec:discussions}

Here we comment on additional effects that may affect our sedimentation 
calculations.  First, turbulent mixing will likely play an important role in 
determining the efficiency of sedimentation.  Turbulence, on the one hand, 
will tend to mix fluid elements from different parts of the cluster and 
to counteract the effect of sedimentation, but it may also enhance the rate of 
sedimentation by increasing the mobility of the ions.  Detailed investigation of 
the effect of turbulence on sedimentation process is out of the scope of the 
present work.   But, we point out that recent hydrodynamical simulations 
of galaxy cluster formation uniformly indicate the presence of ubiquitous subsonic 
turbulent flow in clusters \citep{norman_bryan99,nag03,rasia_etal04,kay_etal04,
faltenbacher_etal05,dolag_etal05,rasia_etal06,2007Nagai}.  
These simulations indicate that the turbulence can provide about 
$5-10$\% of the total pressure support at $r=r_{500}$ even in relaxed systems, 
and its relative importance increases in cluster outskirts as well as unrelaxed 
clusters with recent major mergers (Lau, Kravtsov, Nagai, in preparation).  
It is therefore possible that turbulent mixing caused by gas accretion in 
cluster outskirts and/or major mergers may significantly modify the efficiency 
of the sedimentation.  Hydrodynamical cluster simulations that also include 
the sedimentation physics will likely provide realistic assessment of this effect.

We also point out that the standard diffusion equations assume that 
the mean-free-path of the ions is small compared to the size of the system.  
The proton mean-free-path, however, could become a sizable fraction of 
the cluster size in their outskirts \citep{2007Loeb}.  For a $T_X = 5$~keV 
isothermal cluster, for example, the mean-free-path of proton is comparable 
to the cluster virial radius ($\approx 2 \times r_{500}$).  The validity 
of the fluid approximation, however, depends sensitively on the ICM 
temperature profile.  For the cluster with the same $T_X$, the mean-free-path 
in the V06 model, for example, is considerably smaller (9\%) fraction of the 
virial radius, which makes the fluid approximation still reasonable in the 
radial range of our interests.
 
\begin{figure}[t]
  \centering \includegraphics[clip=true,trim=0.0cm 2.0cm 0.0cm 3.0cm,width=3.8in]{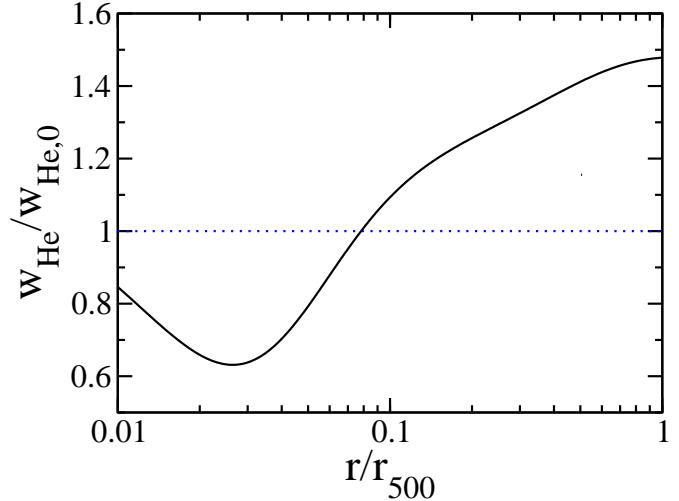}
  \caption{ The ratio of two He sedimentation velocities derived with ($w_{\rm He}$) and 
without ($w_{\rm He, 0}$) the effect of thermal diffusion, plotted as a function of the 
cluster-centric radius. }
  \label{whe.f}
  \vspace{0.2mm}
\end{figure} 

In this work, we also neglected the effect of thermal diffusion, which can cause 
the change in the diffusion speed.  To assess the effect of thermal diffusion, we 
follow the formalisms presented in \citet{burgers69:composite_gases} which takes 
into account the heat flow interferences, and apply it to the H-He 
plasma in the ICM.  For isothermal case, the diffusion velocity is enhanced by 
$\sim 23\%$, which agrees with the value reported by \citet{Chuzhoy2004Element}.  
For non-isothermal case, we find that thermal diffusion enhances He sedimentation 
in the outskirts where the temperature gradient is negative, but suppresses He 
sedimentation in the inner region ($r/r_{500} < 0.1$) with positive temperature 
gradient.   Our analysis indicates that the contribution of thermal diffusion to the 
diffusion velocity by about $<50\%$ throughout clusters.   Figure~\ref{whe.f} shows 
the change in He sedimentation speed caused by 
the effect of thermal diffusion as a function of $r/r_{500}$.   Note that the thermal 
diffusion causes relatively small change in the $Y/X$ at cluster outskirts 
(e.g., $r\sim r_{500}$), despite its large fractional change in diffusion velocity.  
It is because that the diffusion speed was very small there to start with.  Using the 
scaling of $Y/X$ with time given by \citet{Chuzhoy2004Element}, we find that the effect 
of thermal diffusion on $Y/X$ is less than 20\%.
 
It has been suggested that the enhancement of He abundance in the central 
regions might cause the decline of the observed Fe abundance 
\citep{Ettori2006Effects}.  To assess this effect, we repeat the He sedimentation 
calculations for the three cool-core clusters (including Centaurus, A2199, A1795)
presented in \citet{Ettori2006Effects}, using their observed gas density and 
temperature profiles.  Our calculations indicate that the enhancement of 
He in the core of the Centaurus cluster is less than 20\% from the primordial 
He abundance, even in the extreme case of un-magnetized cluster with the 
cluster age of 11 Gyr.  This corresponds to a decline of the observed Fe 
abundance by only $7\%$ \citep[see Fig.~4 of][]{Ettori2006Effects}.  Similar 
results are found for A2199 and A1795.   Note that the He sedimentation is 
significantly suppressed in the cool core regions.  We thus conclude that the 
observed $20-50\%$ reduction in the iron abundance in some of the cool-core
clusters cannot be explained by the He sedimentation alone.

\section{Conclusions}
\label{sec:conclusions}

In this work, we investigate effects of He sedimentation on X-ray measurements 
of galaxy clusters and their implication for cosmological constraints derived
from these observations.  By solving a set of 
flow equations for a H-He plasma and using observationally motivated 
cluster models, 
we show that the efficiency of He sedimentation is significantly suppressed in the
cluster outskirts due to the observed temperature drop, 
while it is dramatically enhanced in the cluster core regions.
Our sedimentation 
model based on the observed temperature profile suggest that the effect of 
helium sedimentation is negligible at $r_{500}$, and it does not affect cluster 
mass measurements obtained at the sufficiently large cluster radius.  
However, the effect of sedimentation increases toward the inner regions 
of clusters and introduces biases in X-ray measurements of galaxy clusters.
For example, at $r_{2500}$, biases in X-ray measurements of gas 
mass, total mass, and gas mass fractions, are at the level of $5-10\%$.  
The effect of He sedimentation could also introduce biases in the estimate 
of the Hubble parameter derived from the combination of X-ray and SZE 
measurements, which could explain the observed offset in the X-ray+SZE 
derived $H_0$ and independent measurement from the Hubble Key project.  
We emphasize, however, that the magnitude of these biases depends 
sensitively on the cluster age, temperature, and magnetic and/or turbulent 
suppression in the ICM.  

We show that the process of He sedimentation introduces the apparent
evolution in the observed gas mass fractions of X-ray luminous,
dynamically relaxed clusters.  The effect of He sedimentation could 
lead to biases in observational constraints of dark energy equation 
of state $w$ at a level of  $\lesssim$10\%.
These biases tend to make the value of $w$ more negative.
Current measurements based on $f\gas$ evolution
\citep{Allen2004Constraints_on_,Allen2008Improved} should not be
significantly affected by these biases.  However, future measurements
aiming to constrain $w$ to better than 10\% may need to take into
account the effect of He sedimentation.  For cosmological
measurements, one way to minimize these biases is to extend the X-ray
measurements to a radius well beyond $r_{2500}$.  At the
same time, the evolution of cluster gas mass fraction in the inner
regions of clusters should provide unique observational diagnostics
of the He sedimentation process in clusters.

\acknowledgements 
We thank Marc Kamionkowski, Avi Loeb, Maxim Markevitch, and 
Sterl Phinney for useful comments on this work.  We also acknowledge
Avi Loeb for suggesting to consider the effect of turbulent mixing and 
the mobility of ions in the ICM.  We thank the anonymous referee for 
helpful comments that greatly improved this work. This work is 
supported by Sherman Fairchild Foundation.

\bibliographystyle{apj}
\bibliography{ms}

\end{document}